\documentclass[twocolumn,trackchanges]{aastex631}

\usepackage{wrapfig}
\usepackage{float}
\usepackage{placeins}
\usepackage{multirow}
\usepackage{mwe}
\usepackage{hyperref}
\usepackage{amsmath}

\newcommand{\cms}{{\rm cm}^{-2}}

\newcommand{\msolar}{{\rm M}_{\odot}}

\newcommand{\HI}{{\hbox{H\,{\sc i}}}}

\newcommand{\OVII}{\hbox{O\,{\sc vii}}}
\newcommand{\OVIII}{\hbox{O\,{\sc viii}}}
\newcommand{\FeXVII}{\hbox{Fe\,{\sc xvii}}}
\newcommand{\LEM}{$\mu$cal}
\newcommand{\fgas}{f_{\rm gas}}
\newcommand{\fcool}{f_{\rm cool}}
\newcommand{\HItab}{H\,{\sc i}}

\newcommand{\sechead}[1]{%
  \multicolumn{10}{|c|}{\rule[-0.9ex]{0pt}{3.4ex}\boldmath\bfseries #1}}

\begin{document}

\title{CNN-Based Inference of Gaseous Halo Properties from Synthetic X-ray and 21-cm HI Observations\footnote{Submitted September, 2026}}

\author[0009-0005-7596-4204]{Kalvyn N. Poncelet Adams}
\thanks{Email: adams.kalvyn@gmail.com}
\affiliation{University of California, Los Angeles, Department of Physics and Astronomy, Box 951547, Los Angeles, CA 90095, USA}

\author[0000-0002-3391-2116]{Benjamin D. Oppenheimer}
\affiliation{University of Colorado, Center for Astrophysics and Space Astronomy, 389 UCB, Boulder, CO 80309, USA}

\author[0009-0003-4100-4710]{Naomi Gluck}
\affiliation{Department of Physics, Yale University, New Haven, CT 06520, USA}

\author[0009-0003-9247-8617]{Caleb Ogle}
\affiliation{Department of Physics and Astronomy, University of Wisconsin-Milwaukee, Milwaukee, WI 53201, USA}

\author[0000-0003-3207-8868]{Matthew Ho}
\affiliation{Sorbonne Universit\'{e}, CNRS, UMR 7095, Institut d'Astrophysique de Paris, 98 bis bd Arago, 75014 Paris, France\\}
\affiliation{Department of Astronomy, Columbia University, New York, NY, 10027}

\author[0000-0002-6766-5942]{Daisuke Nagai}
\affiliation{Department of Physics, Yale University, New Haven, CT 06520, USA}
\begin{abstract}

Quantifying the information content in multi-wavelength observations is critical for setting exposure times for upcoming X-ray and 21-cm \HI\ radio surveys. We train convolutional neural networks (CNNs) on mock observations of halos from the IllustrisTNG100 and TNG300 simulations, combining data from soft X-ray channels from a CCD or a microcalorimeter with \HI\ intensity, velocity, and dispersion maps, to infer halo mass, gas fractions, metallicity, and [O/Fe] abundance. Multi-band (X-ray and \HI) combinations consistently outperform single-band inference for gas fractions. X-ray outperforms \HI\ observations for measuring halo mass, but both bands contribute similarly when measuring the cool ($T<10^5$ K) gas fraction in halos with significant cool gas content.  Using matched exposure times, a micro-calorimeter improves metallicity inference over the CCD by a factor of 1.75, enabling precise measurements of [O/Fe] alpha-enhancement for the largest halos.  The larger volume of TNG300 allows inference of group halo masses, finding an inference RMSE of 0.04 dex with a 100 ksec X-ray exposure time. These results demonstrate how deep learning can evaluate strategies for developing instruments and designing surveys for these expensive observations targeting gaseous halos.  

\end{abstract}


\section{Introduction} \label{sec:intro}

As astronomical observations continue to move to larger and more expensive observational facilities, a primary goal for current and future telescopes is deep, targeted observations of carefully selected objects. Therefore, it is crucial to quantify the information content obtained from these observations, including efficiencies of instruments, effective exposure times, and the synergy of facilities. We can evaluate mock observations created by simulations with deep learning (DL) to determine the scientific value of various observational survey specifications and whether they meet the requirements for achieving precise results. 

The circum-galactic medium (CGM) is the diffuse, multi-phase gas residing between a galaxy's interstellar medium and its virial radius. Studying the CGM is of paramount importance for revealing the gaseous supply necessary to feed star formation \citep[e.g][]{prochaska_nonevolution_2009, christensen_-n-out_2016, angles-alcazar_cosmic_2017}, mapping the distribution of metal products released by stellar nucleosynthesis and feedback \citep[e.g.][]{peeples_budget_2014, Lovisari2019}, and determining the process by which galaxies quench \citep[e.g.][]{tumlinson_large_2011,Zhang2025}. Thus, mapping the multi-phase CGM requires developing systematic methods ensuring that future surveys are built to maximize information content for deep, targeted observations.  

The multi-phase CGM can be broken up into two thermodynamic phases: the cool phase ($T\sim10^4$ K) and the warm-hot phase ($T> 10^{5-6}$ K). The cool gas is most often measured along targeted sight lines toward quasars using UV absorption line measurements \citep{tumlinson_cos-halos_2013, werk_cos-halos_2013, Stocke2013,chen_characterizing_2018}. It can also be observed in high-resolution 21-cm emission via radio interferometric arrays to map the cool phase. 21-cm $\HI$ is well mapped within the interstellar medium (ISM) by existing radio telescopes \citep[e.g.]{Oosterloo2010,Heald2011}, but far deeper observations are necessary for CGM $\HI$ due to lower gas densities.  The observational campaign of MHONGOOSE \citep{De_Blok2024-hi} using the MeerKAT telescope and the WALLABY observation of group NGC 7162 \citep{reynolds_wallaby_2019} with the ASKAP telescope are examples of these deeper integrations required to measure the CGM.  

The warm-hot phase is most often observed using deep X-ray observations \citep[e.g.][]{bogdan_correlation_2018, bregman_extended_2018, mathur_probing_2023}. Soft X-ray emission is detectable from galactic sources in typical galaxies \citep[e.g.][]{Li2013a,Goulding2016}. X-ray imagers include CCD detectors, such as  {\it Chandra} ACIS \citep{10.1117/12.3018498}, {\it XMM-Newton} EPIC, and {\it eROSITA} \citep{predehl_erosita_2021}. There are also microcalorimeter cameras, including the ongoing {\it XRISM} \citep{2020arXiv200304962X} as well as future prospective cameras, including {\it Athena} X-IFU \citep{barret_athena_2016} and {\it Explorer of Cosmic Ecosystems and Energetic Dynamics} ({\it ExCEED}), a $\sim 25\%$ smaller version of the {\it Line Emission Mapper} \citep[\textit{LEM},][]{Kraft2003}. Microcalorimeter-enabled spectral resolution can image and disentangle individual emission lines, including $\OVII$, $\OVIII$, and $\FeXVII$, from other backgrounds.

Even with the vast array of surveys, more expensive observations are necessary to reliably detect extended emission from the hot phase CGM \citep[e.g.][]{anderson_deep_2016, bogdan_correlation_2018, Das2020}. For example, the sensitivity of {\it XMM-Newton} EPIC declines significantly beyond $\sim 40\ {\rm kpc}$, while the extended emission from filaments of NGC 3079 extend to $\sim60\ {\rm kpc}$ from the galactic center \citep{HodgesKluck2020}. Thus, the next step in mapping the hot CGM is to devise the most efficient survey configuration that maximizes information content. The information content of the simulated X-ray and $\HI$ CGM observations has been explored using convolutional neural networks (CNNs) applied to \texttt{CAMELS} \citep[Cosmology \& Astrophysics MachinE Learning Simulations][]{camels_2021} in \citet{gluck_observationally_2024}, hereafter G24. G24 focused on the ability to measure a variety of CGM properties, including halo mass, CGM mass, and CGM metallicity from single-band X-ray combined with \HI\ column density. A key result from G24 is that the combination of X-ray and radio mapping can outperform the inference of underlying CGM properties better than a single band on its own.  

Extending on the work of G24, we use CNNs to study the information content available from multiple X-ray bands (narrow and wide via a microcalorimeter) and 21-cm measurements of velocity and dispersion. As the small box size of \texttt{CAMELS} limited the ability to infer CGM properties of larger objects ($M>10^{14.0}\ M_{\odot}$), we use the \texttt{IllustrisTNG100} and \texttt{IllustrisTNG300} simulations \citep[][hereafter, TNG100 and TNG300, respectively]{pillepich_simulating_2018,springel_first_2018,Federico_2018, Naiman_2018,Nelson_illustristng_2018}.

We train our CNN on various combinations of wavelength bands from TNG100 or TNG300. We then test the trained CNN on an unseen portion of the same dataset. We specifically want to determine 1) whether the high exposure time used to obtain X-ray data is worth it given the information gained, 2) how to demonstrate the advantages of a high spectral resolution microcalorimeter over a standard CCD camera quantitatively for X-ray detections, and 3) the information increase one obtains when utilizing $\HI$ velocity maps. 

This paper is outlined as follows. Section~\ref{sec:methods} explains the specific simulation setup, the generation of mock observations, the design and architecture of the CNN, and the output and analysis of the results. Our results are presented in Section~\ref{sec:results}: Section~\ref{sec:mass_results} includes using TNG100 to infer mass quantities and radio with X-ray, \HI, and \HI\ moment maps; Section~\ref{sec:Z_results} includes using TNG100 to infer metallicity and abundance ratios with X-ray; Section~\ref{sec:TNG300_M500c} includes using TNG300 to infer halo mass of grup-sized halos with all available channels. Section~\ref{sec:discuss} discusses the performance of our network and compares it with previous results. Section~\ref{sec:summary} concludes. 

\section{Methods} \label{sec:methods}

We begin this section with a description of the TNG100 and TNG300 simulations from which we derive the mock observations (\S~\ref{2.1 Simulations}). We then discuss how we create the mock observations and galactic parameters used to train the network (\S~\ref{2.2 Data Generation}), the design of the Convolutional Neural Network (\S~\ref{2.3 CNN}), and the output of the CNN and how we measure its performance (\S~\ref{sec:metrics}).  

\subsection{Simulations} \label{2.1 Simulations}  

The publicly available TNG100 and TNG300 \citep{Naiman_2018, marinacci_2018, pillepich_simulating_2018, springel_first_2018, Nelson_illustristng_2018} cosmological hydrodynamic simulations use the AREPO hydrodynamical scheme \citep{springel_2010}. AREPO uses an N-body tree-particle-mesh method to integrate gravity and a moving mesh for magnetohydrodynamics. The IllustrisTNG subgrid physics modules include star formation and evolution, stellar-driven superwinds, black hole accretion and merger, and black hole feedback. Specifically, black hole feedback is implemented in a dual-mode approach, where thermal feedback is implemented in the case of high Eddington accretion rates, and kinetic feedback is used in the low Eddington accretion rate regime \citep{Weinberger2017, pillepich_simulating_2018}. The simulation is evolved to $z=0.00$, and we use the final snapshot for the generation of our mock data.

We use TNG100 for its ability to resolve the CGM around typical galaxies. TNG300 offers a larger sample of groups and cluster halos, which is necessary to sufficiently train a CNN, but at lower resolution $\sim 8\times$. Table~\ref{tab:TNG_values} highlights specific differences between the two simulations. For additional details, refer to the IllustrisTNG documentation\footnote{ \href{https://www.tng-project.org/about/}{https://www.tng-project.org/about/}}.

\begin{table}
    \centering
    \caption{IllustrisTNG simulation values for TNG100 and TNG300 via  \citet{nelson_illustristng_2019}. }  
    \label{tab:TNG_values}
    \begin{tabular*}{\columnwidth}{|m{3cm}|m{2.2cm}|m{2.2cm}|}
    \hline
    & \textbf{TNG100}  & \textbf{TNG300} \\
    \hline
    Volume [Mpc$^3$] & $110.7^3$ & $302.6^3$\\
    Length [Mpc/$h$] & 75 & 205 \\
    $n_{\rm DM}$ & $1820^3$ & $2500^3$\\
    $n_{\rm gas}$ & $1820^3$ & $2500^3$\\
    $m_{\rm DM}$ [$M_\odot$]& $7.5\times10^6$ & $5.9\times10^7$\\
    $m_{\rm gas}$ [$M_\odot$] & $1.4\times10^6$ & $1.1\times10^7$\\

    \hline
    \end{tabular*}
    
\end{table}

\begin{figure*}

\includegraphics[width=0.195\textwidth]{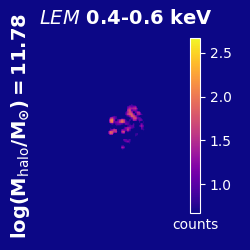}
\includegraphics[width=0.195\textwidth]{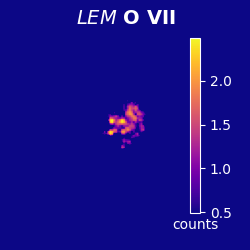}
\includegraphics[width=0.195\textwidth]{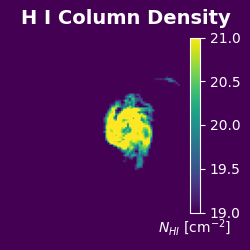}
\includegraphics[width=0.195\textwidth]{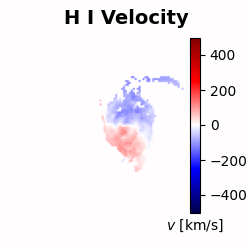}
\includegraphics[width=0.195\textwidth]{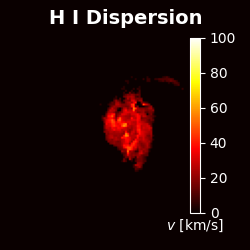}

\includegraphics[width=0.195\textwidth]{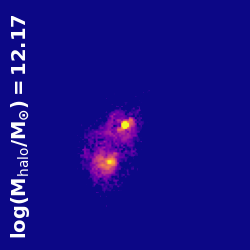}
\includegraphics[width=0.195\textwidth]{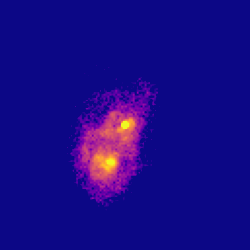}
\includegraphics[width=0.195\textwidth]{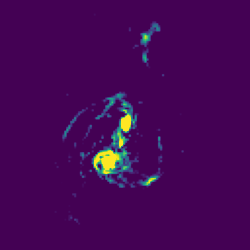}
\includegraphics[width=0.195\textwidth]{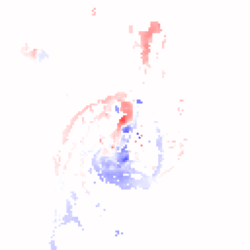}
\includegraphics[width=0.195\textwidth]{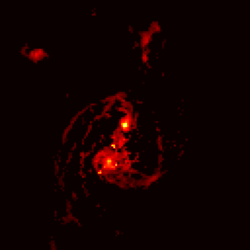}

\includegraphics[width=0.195\textwidth]{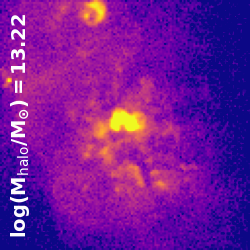}
\includegraphics[width=0.195\textwidth]{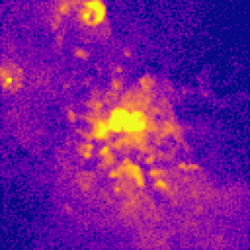}
\includegraphics[width=0.195\textwidth]{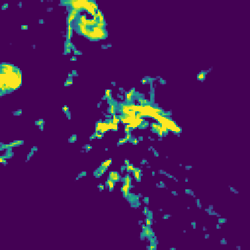}
\includegraphics[width=0.195\textwidth]{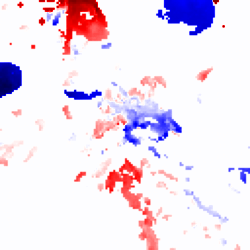}
\includegraphics[width=0.195\textwidth]{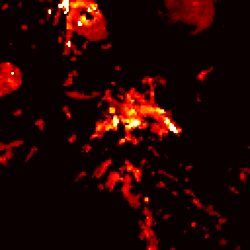}

\includegraphics[width=0.195\textwidth]{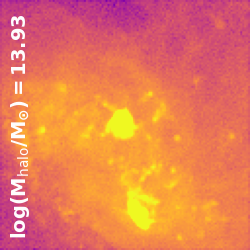}
\includegraphics[width=0.195\textwidth]{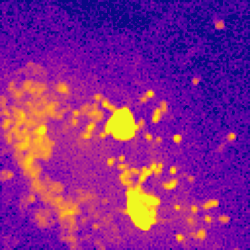}
\includegraphics[width=0.195\textwidth]{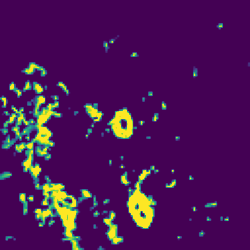}
\includegraphics[width=0.195\textwidth]{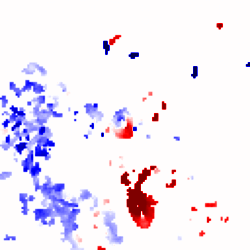}
\includegraphics[width=0.195\textwidth]{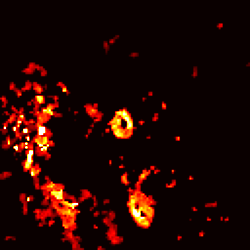}
\caption{Halo maps from TNG100 input into our CNN. Four halo masses are shown increasing from top to bottom. From left to right, the {\it Line Emission Mapper} ({\it LEM}) mocks made via pyXSIM wide-band 0.4-0.6 keV channel, the narrow-band \OVII\ channel (0.57 \AA), the \HI\ radio 21-cm spin-flip channel, the \HI\ velocity shift, and the \HI\ velocity dispersion.}
\label{fig:snaps}
\end{figure*}

\subsection{Mock Data Generation} \label{2.2 Data Generation}

Galaxy and group halos are selected from \texttt{SUBFIND} group catalogues \citep{springel_subfind_2001} downloadable from the IllustrisTNG website.  Our data sample consists of 3566 galaxy halos with masses between $M_{500}=10^{11.7}-10^{14.0}\;\msolar$, where $M_{500}$ is the total mass within a sphere of $500\times$ the critical overdensity returned by \texttt{SUBFIND}. We additionally calculate the total gas fraction,
\begin{equation}\label{eq:fgas}
f_{\rm gas}\equiv \frac{M_{\rm gas}(<R_{500})}{M_{500}},
\end{equation}
and the cool gas fraction,
\begin{equation}\label{eq:fcool}
f_{\rm cool}\equiv \frac{M_{\rm gas}(T<10^5 K,<R_{500})}{M_{\rm gas}(<R_{500})}.
\end{equation}
We generate neutral hydrogen (\HI) and X-ray maps using \texttt{yT} Project \citep{turk_yt_2011} software as described in \S~\ref{sec:soft_xray_generation} and \S~\ref{sec:HI_generation}. 
\subsubsection{Soft X-ray}
\label{sec:soft_xray_generation}

We develop our pipeline to create X-ray mocks for both a traditional CCD detector and a high-spectral-resolution microcalorimeter (hereafter, \LEM) detector. Our \LEM\ detector setup adopts the proposed NASA mission {\it Line Emission Mapper} parameters \citep{Kraft2003}, which have since been adapted into the {\it ExCEED} mission concept \citep{Kraft2026}. \textit{ExCEED} features a similar detector, but utilizes a slightly smaller mirror layout, meaning that our mocks remain highly representative but require longer integration times ($\sim 35\%$) in practice. The \LEM\ instrument has high energy resolution, allowing the direct spatial mapping of individual emission lines. Similar \LEM\ mocks have been extensively explored in a series of recent papers \citep{truong_x-ray_2023, nelson_resonant_2023, schellenberger_mapping_2024, zuhone_properties_2024}. Our pipeline configurations are also broadly applicable to other future microcalorimeter instruments, including {\it ATHENA}'s X-IFU.  

The pipeline begins by using \texttt{pyXSIM} \citep{zuhone_pyxsim_2016} to calculate the X-ray photon list (saved as a ``SIMPUT'' file) from the simulated fluid gas elements. We then generate mock observations using the \texttt{SOXS} package \citep{zuhone_soxs_2023} based on the instrumental parameters of {\it \LEM}, assuming a spectral energy resolution of 0.9 electron-volts (eV). The detector geometry is configured with a 15-arcsecond pixel size across a 128-by-128 pixel grid, yielding a total field of view (fov) of 32 arcminutes (or approximately 0.53 degrees) on a side. The mocks are projected at a redshift of $z=0.01$ and incorporate realistic instrumental backgrounds and Galactic foregrounds, which we subtract during post-processing. For simplicity, we omit the Cosmic X-ray Background, as it can be removed via established empirical algorithms \citep[e.g.][]{oppenheimer_eagle_2020}. 

To isolate specific physical features, we apply dynamic filters adjusted for the peculiar velocity of each simulated galaxy. This produces 7 total energy bands: 4 continuum-dominated widebands (0.2-0.4 keV, 0.4-0.6 keV, 0.73-1.1 keV, and 1.43-2.0 keV), and 3 line-dominated narrowbands targeting $\OVII$, $\OVIII$, and $\FeXVII$. TNG100-based LEM mock-maps of the 0.4-0.6 keV wideband and $\OVII$ are in the first and second columns from the left in Figure~\ref{fig:snaps}, respectively. 

Except for the metal-rich 0.73-1.1 keV band, the widebands intentionally exclude major emission lines to isolate true continuum emission. The $\OVII$ narrowband incorporates three distinct sub-bands spanning 0.561-0.574 keV to fully capture this characteristic triplet. The $\OVIII$ filter isolates the single resonance line at 0.654 keV, while the $\FeXVII$ filter covers four prominent lines ranging between 0.725 and 0.826 keV.

For comparison, we generate an equivalent set of CCD-like observations. This setup uses the same 15-arcsecond pixel size and 32-arcminute FOV, but is restricted to the 4 widebands (0.2-0.4 keV, 0.4-0.6 keV, 0.7-1.1 keV, and 1.4-2.0 keV) without narrowband or sub-band capability, masking out major metal-line channels. This baseline configuration approximates the capabilities of current-generation imaging X-ray spectrometers. It closely mirrors deep observations from the \textit{XMM-Newton} EPIC cameras (30-arcminute FOV and 15-arcsecond spatial resolution), and the \textit{Chandra} ACIS-I observations when spatially binned to the same 15-arcsecond scale. This dual-detector approach allows us to directly quantify the diagnostic advantages of future \LEM-level resolution over traditional CCD architecture. We assume deep exposure timescales of 1 megasecond (Msec) for the TNG100 volume, and 100 kiloseconds (ksec) for the TNG300 volume.

\subsubsection{Neutral Hydrogen}
\label{sec:HI_generation}

We generate idealized $\HI$ column-density maps as our primary observational signal. This noise-free framework is used to evaluate gas kinematics at a uniform scale, instead of modeling a specific single-dish telescope or interferometer layout. 

The \texttt{Trident} module \citep{hummels_trident_2017} is used to generate \HI\ fractions given the strength of the ionizing background \citep{haardt_radiative_2012} and self-shielding \citep{rahmati_impact_2013}. 

To ensure direct spatial alignment and facilitate cross-correlation with our X-ray mocked data, we impose the same grid geometry: 15-arcsecond pixels across a 128-by-128 pixel field of view (spanning 32 arcminutes on a side). At our sample redshift of $z=0.01$, this angular resolution corresponds to a physical pixel size of approximately $3.2$ kiloparsecs (kpc) and a total field width of approximately 407 kpc. This configuration directly maps to the observational capabilities of the \textit{MeerKAT} radio interferometer array configured for the MHONGOOSE survey \citep{De_Blok2024-hi}. A column density detection limit of $10^{19}\;\cms$ is used based on observations by the design of the MHONGOOSE Survey, following initial results \citep{healy_possible_2024}.

Using the particle-particle-velocity cube (PPVCube) function within \texttt{yT} \citep{turk_yt_2011}, we construct 3D data cubes featuring a kinematic channel resolution of 10 ${\rm km}\ {\rm s}^{-1}$. Mock velocity (first-moment) and velocity dispersion (second-moment) maps are derived by calculating the column density-weighted mean and variance for all voxels exceeding a local limit of $10^{18}\ {\rm cm^{-2}}$. The three right-most panels of Figure~\ref{fig:snaps} show these structural variations across our halo mass range using the three primary \HI\ data products: the 0th-moment column density, the 1st-moment velocity, and the 2nd-moment dispersion maps. This framework remains consistent with G24, which used identical column-density bounds to evaluate the 21-cm \HI\ 0th-moment projections.    

\subsection{Convolutional Neural Network}\label{2.3 CNN}

We use a Convolutional Neural Network (CNN) as a likelihood-free inference framework \citep{cranmer2020frontier} to quantify the information gain from including uncorrelated gas-structure channels. Traditional analytical modeling often struggles because the brightness and shape of the \HI\ and the corresponding velocity and dispersion maps are not linearly related to galaxy mass \citep{Davies_2011}. Because CNNs can learn features across multiple fields simultaneously, they can accept both intensity maps and non-intensity kinematic data, such as \HI\ velocity and dispersion. This capacity to analyze both magnitude and structure yields more accurate estimates of the halo mass and other CGM properties.

The CNN is built using PyTorch \citep{2019arXiv191201703P} and is highly malleable in two key respects: the input stage dynamically adapts to the number of input channels, allowing the network to analyse up to 14 different wavelength mappings per galaxy, and the target parameters are fully configurable, allowing certain runs to focus on mass and gas fractions while others focus on metallicity and [O/Fe]. This flexibility allows systematic comparison of individual fields against multi-wavelength combinations, and ensures that easily estimated parameters are not mixed with more challenging ones. Hyperparameters (learning rate, dropout, weight decay, and number of hidden layers) are optimised using \texttt{Optuna} \citep{akiba_2019}, with best-performing configurations consistently found near a learning rate $\sim 4\times10^{-4}$, dropout $\sim 0.05$, weight decay $\sim 0.01$, and $5$ hidden layers. We note that the \LEM\ network favors a larger weight decay compared to \LEM-\HI\ and CCD-\HI. Full implementation details are available at \href{https://github.com/IStoleTheCookieJar/DLHD-CNN}{https://github.com/IStoleTheCookieJar/DLHD-CNN}.

The forward pass produces an output tensor $\mathbf{\hat{\Theta}} \in \mathbb{R}^{N \times 2}$, where each row $i \in \{1, \dots, N\}$ yields a vector $v_i=(\mu_i, \sigma_i)$, with $\mu_i$ the parameter estimate and $\sigma_i$ the input-dependent uncertainty. Training proceeds in two stages with different loss functions. During pre-training, we use mean squared error (MSE):
\begin{equation}\label{eq:MSE}
    \mathcal{L}_{\text{MSE}}(\textbf{y},\boldsymbol{\mu}) = 
    \sum_{i=1}^{N}(y_i - \mu_i)^2.
\end{equation}
During training, we switch to a negative log-likelihood (NLL) loss:
\begin{equation}\label{eg:NLL}
    \mathcal{L}_{\text{NL}}(\textbf{y},\boldsymbol{\mu},\boldsymbol{\sigma}) = 
    -\sum_i \left(-\log(\sigma_i) - \frac{1}{2}\left(\frac{y_i - \mu_i}
    {\sigma_i + \epsilon} \right)^2 \right),
\end{equation}
where $\epsilon=10^{-7}$ prevents divergence at low uncertainties. For further justification of these loss functions, see Ogle et al. (in prep).

We split the full dataset into 64\%, 16\%, and 20\% for training, validation, and testing, respectively. Following G24, different projection axes (x, y, z) of the same galaxy are barred from appearing in both the training and testing sets to prevent memorization within the network.

\subsection{Network Output Analysis}\label{sec:metrics}

We output several metrics, similar to G24 and \citet{10.1093/mnras/staf1888}.  For each true value of a property $y_i$ for an instance $i$, across a sample of $N$ instances, the CNN infers predictions $\mu_i$ with errors $\sigma_i$.  

The first metric is the root mean squared error (RMSE), which determines the accuracy of the model in units of the measured property, as:
\begin{equation}
        \rm{RMSE} = \sqrt{\langle (y_i - \mu_i)^2\rangle},
        \label{eqn:rmse}
\end{equation}
The second metric is the Pearson correlation coefficient, $\rho$, which represents the linear correlation between the true and predicted parameter values:
\begin{equation}
        \rho = \frac{\Sigma_i (\mu_i-\overline{\mu})(y_i - \overline{y})}{\sqrt{\Sigma_i (\mu_i - \overline{\mu})^2}\sqrt{\Sigma_i (y_i - \overline{y})^2}},
    \label{eqn:r_squared}
\end{equation}
where $\overline{y}$ is the mean of the inputs and $\overline{\mu}$ is the mean of the predictions. If $\rho=1$, there is a perfectly linear correlation; $\rho=0$ represents zero correlation, and $\rho=-1$ represents perfect anti-correlation. 
The last metric, the reduced chi-squared, $\chi^2$, quantifies the ``trustworthiness'' of the posterior standard deviation, defined as:
\begin{equation}
        \chi^2 = \frac{1}{N} \sum_{i=1}^{N} \left(\frac{y_i - \mu_i}{\sigma_i} \right)^2.
        \label{eqn:chi2}
\end{equation}
Values close to one indicate a properly quantified error, and values less than (greater than) one overestimate (underestimate) error. We omit the mean relative error.

\section{Results} \label{sec:results}

We discuss our results in the following sequence.  We begin by examining how the mass quantities ($M_{500}$) and ratios ($\fgas$ and $\fcool$) are inferred using $\HI$ and two different types of X-ray observations (CCD and \LEM), focusing on galaxies in \S\ref{sec:mass_with_xray_HI}.  We then use \HI\ moment maps to infer the same properties, showing how the addition of velocity and dispersion can better determine mass measurements in \S\ref{sec:moment_results}.  We then demonstrate the results of adding the X-ray resolution of a \LEM\ to determine the CGM metallicity and abundance ratios in \S\ref{sec:Z_results}.  These first three sections use the TNG100 simulation, which lacks a sufficient sample of group-mass halos for training.  We therefore infer group and cluster properties using the larger volume TNG300 simulation in \S\ref{sec:TNG300_M500c}.

\subsection{Mass Quantities and Ratios}
\label{sec:mass_results}

\subsubsection{Inference with X-ray and Neutral Hydrogen}
\label{sec:mass_with_xray_HI}

\begin{figure*}
\includegraphics[width=\textwidth]{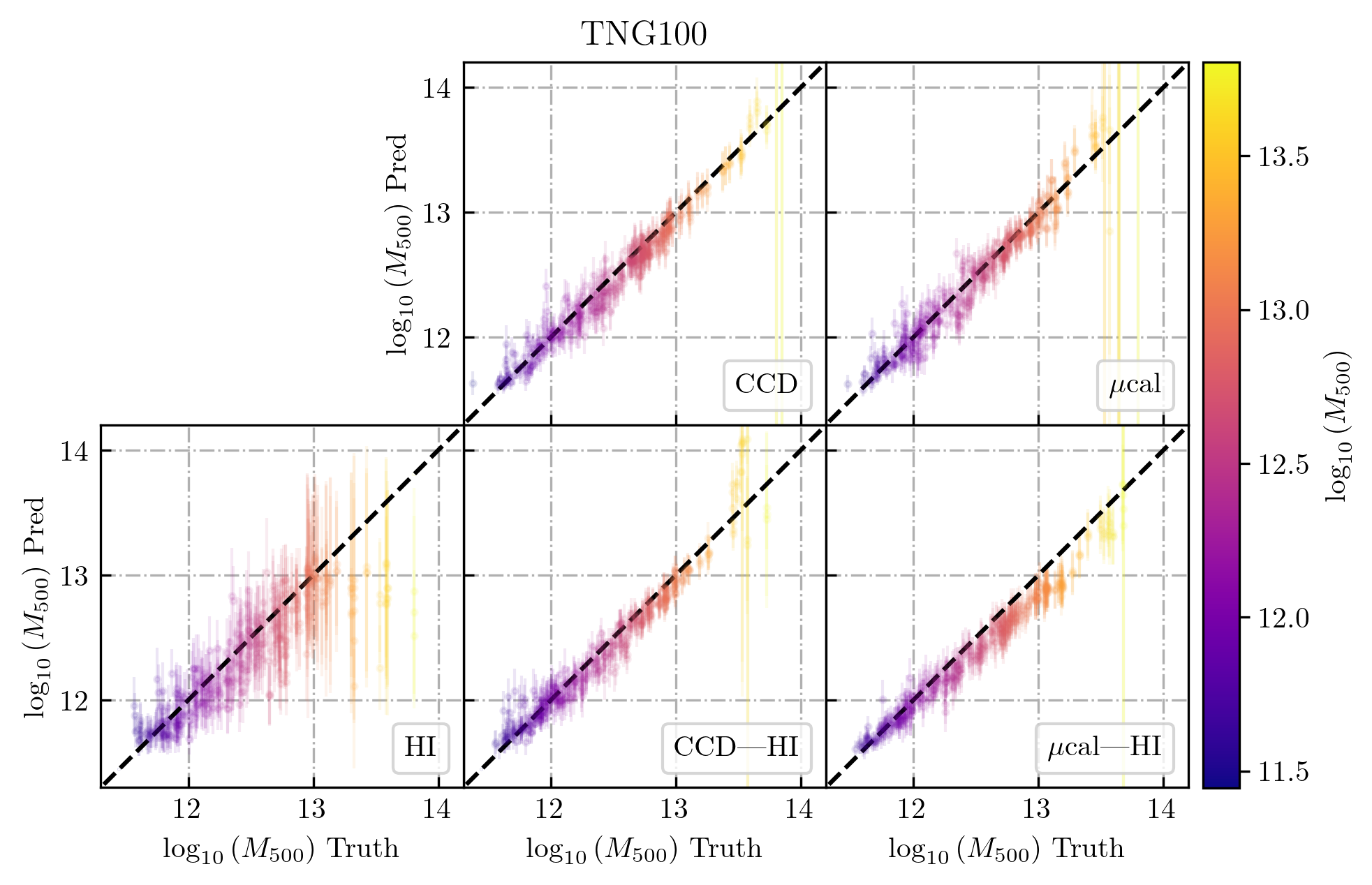}
\caption{Truth-inference plots for $M_{500}$ in TNG100 using different combinations of observations. \textit{Top Row:} Results using X-Ray CCD, and separately using \LEM. \textit{Bottom Row:} Results using \HI\ separately, X-ray CCD with \HI, and \LEM\ with \HI. The black, dashed line indicates perfect inference. The points are colored by their true $M_{500}$ value. For readability, we plot a mass-dependent fractional sample of halos from the test set, decreasing the number of low-mass halos plotted while keeping most high-mass halos in the figure. The corresponding statistics for the full test set are in Table~\ref{tab:TNG100_M500}.}
\label{fig:TNG100_M500c}
\end{figure*}

Figure~\ref{fig:TNG100_M500c} shows the CNN-inferred halo mass, $M_{500}$, of TNG100. We plot inferred values against the true values from the held-out test set in {\it Truth-Inference} panels for CNN runs with different input channel combinations. Our layout shows a logical progression of channel combinations, with more channels proceeding from left to right and the addition of the \HI\ column density channel along the bottom.  The upper panels show X-ray CCD results using 4 channels (upper center) and X-ray \LEM\ results using 7 channels (upper right).  The lower panels begin on the left with \HI\ single channel only and proceed to the right with multiband 5-channel CCD-\HI\ and 8-channel \LEM-$\HI$.

\begin{center}
\begin{table*}[t]
	\centering
	\caption{The inference statistics (root mean square error, Pearson correlation coefficient, and reduced chi-squared) for TNG100 results,  divided into $M_{500}$ bins. Underneath the inferred properties ($M_{500}$, $f_{\rm gas}$, $f_{\rm cool}$, $\log(Z)$, and $[{\rm O/Fe}]$), the channels used as inputs to the CNN are listed. The results of the first 5 channels (\HI, CCD, CCD-\HI, \LEM, and \LEM-\HI) are grouped in Figs. \ref{fig:TNG100_M500c} ($M_{500}$), \ref{fig:TNG100_f_gas} ($f_{\rm gas}$), and \ref{fig:TNG100_f_cool} ($f_{\rm cool}$). The results for the same properties, but using the \HI, velocity, and dispersion channels, are in Fig.~\ref{fig:TNG100_3_moments}. Inferred $\log(Z)$ and $[{\rm O/Fe}]$ are plotted separately in Fig.~\ref{fig:TNG100_Z_OFe}. Bold values indicate the best value in each column for each property.}
    \begin{tabular*}{13.1cm}{|m{1.8cm}|m{0.9cm}|m{0.7cm}|m{0.7cm}|m{1.0cm}|m{1.0cm}|m{1.0cm}|m{0.9cm}|m{0.7cm}|m{0.7cm}|}
	\hline
	\textbf{Channels} & \multicolumn{3}{c|}{\textbf{$<10^{12.3}$} ($L^\star$)}  & \multicolumn{3}{c|}{\textbf{$10^{12.3}-10^{13.0}$} (Super-$L^\star$)} & \multicolumn{3}{c|}{\textbf{$>10^{13.0}$} (Group)}  \\
	\cline{2-10} & RMSE & $\rho$ & $\chi^2$ & RMSE & $\rho$ & $\chi^2$ & RMSE & $\rho$ & $\chi^2$ \\
	\hline
    \sechead{$M_{500}$}\\
    \hline
	\HItab & 0.15 & 0.69 & 0.64 & 0.21 &  0.68 & 0.66 & 0.54 & -0.3 & \textbf{1.2} \\
	CCD & 0.094 & 0.88 & 0.81 & \textbf{0.11} & 0.92 & 0.66 & \textbf{0.085} & \textbf{0.97} & 0.30 \\
	CCD-\HI & 0.094 & 0.88 & 0.60 & 0.12 &  0.91 & 0.80 & 0.15 & 0.84 & 0.65 \\
	$\mu$cal & 0.10 & 0.85 & \textbf{1.0} & \textbf{0.11} & 0.88 & 0.66 & 0.16 &  0.80 & 0.54 \\
	$\mu$cal-\HI & \textbf{0.082} & \textbf{0.91} & 0.59 &  0.14 & \textbf{0.94} & \textbf{1.02} & 0.21 & 0.94 & 1.5 \\
    \HItab,vel,disp & 0.097 &  0.86 & 0.69 & 0.20 & 0.81 & 1.78 & 0.30 & 0.54 & \textbf{1.2} \\
	\hline
    \sechead{$\fgas$}\\
    \hline
    \HItab & 0.017 & 0.89 & 0.74 & 0.015 & 0.73 & 0.63 & 0.024 & 0.30 & 0.72 \\
	CCD & 0.016 & 0.91 & 1.06 & 0.007 &  0.91 & 0.63 & 0.007 & \textbf{0.94} & 0.51 \\
	CCD-\HI & 0.012 & 0.95 & 0.83 & \textbf{0.006} & \textbf{0.95} & 0.38 & 0.011 & 0.88 &  0.46 \\
	$\mu$cal & 0.015 & 0.92 & 1.05 & 0.009 & 0.91 & 1.19 & \textbf{0.009} & 0.88 & 0.64 \\
	$\mu$cal-\HI & \textbf{0.011} & \textbf{0.96} & 0.66 & \textbf{0.006} & \textbf{0.95} & 0.35 & \textbf{0.009} & 0.90 & 0.59 \\
    \HItab,vel,disp & 0.013 & 0.94 & \textbf{1.04} &  0.010 & 0.89 & \textbf{0.86} & 0.016 & 0.62 & \textbf{1.1} \\
	\hline
    \sechead{$\fcool$}\\
    \hline
    \HItab & 0.096 & 0.88 & 0.75 & 0.11 & 0.86 & 0.79 & 0.11 &  0.49 & 0.38 \\
	CCD &  0.12 & 0.80 & 0.91 & 0.11 & 0.80 & 0.88 & 0.048 & \textbf{0.87} & 0.34 \\
	CCD-\HI & 0.069 & 0.94 & 0.81 & \textbf{0.058} & \textbf{0.96} & 0.63 & \textbf{0.042} & 0.73 & 0.35 \\
	$\mu$cal & 0.11 & 0.84 & \textbf{0.95} & 0.11 & 0.86 & \textbf{1.05} & 0.045 & 0.62 & 0.25 \\
	$\mu$cal-\HI & \textbf{0.065} & \textbf{0.95} & 0.68 & 0.075 & \textbf{0.96} &  0.95 & 0.086 & 0.80 & \textbf{1.3} \\
    \HItab,vel,disp & 0.081 & 0.92 & 0.92 & 0.083 & 0.90 & 0.80 & 0.067 & 0.57 & 0.28 \\
	\hline
    \sechead{$\log(Z)$}\\
	\hline
	CCD & 0.11 & 0.66 & 0.75 &  0.15 & 0.68 & \textbf{0.95} & 0.16 & 0.63 & \textbf{0.87} \\
	$\mu$cal & \textbf{0.095} & \textbf{0.79} & \textbf{0.96} & \textbf{0.085} & \textbf{0.92} & 0.76 & 0.095 & \textbf{0.83} & 0.72 \\
	\hline
    \sechead{[O/Fe]}\\
	\hline
	CCD & 0.052 & 0.92 & 0.80 & 0.059 & 0.85 & \textbf{0.79} & \textbf{0.047} & 0.72 & 0.48 \\
	$\mu$cal & \textbf{0.045} & \textbf{0.95} & \textbf{0.81} &  \textbf{0.040} &  \textbf{0.95} & 0.74 & 0.049 & \textbf{0.84} & \textbf{0.97} \\
    \hline
	\end{tabular*}
\label{tab:TNG100_M500}
\end{table*}
\end{center}

As expected, inference improves as we include more channels, as shown by smaller error bars and a tighter relationship, especially from \HI\ to CCD-\HI.  We quantify the aforementioned statistics by halo mass bin in Table~\ref{tab:TNG100_M500}, with bin labels as: $L^\star$ ($M_{500}= 10^{11.7-12.3}\;\msolar$), super-$L^\star$ ($M_{500}= 10^{12.3-13.0}\;\msolar$), and groups ($M_{500}= 10^{13.0-14.0}\;\msolar$). 

$\HI$ alone exhibits the poorest inference capability by far, likely owing to tracing a decreasing fraction of the gas at higher mass, as well as its non-monotonic relationship with halo mass (see G24, Figure 4).  In contrast, the long X-ray exposure times (1 Msec) show that the halo mass can be powerfully constrained down to the lowest masses.  Furthermore, the best performance is the multi-wavelength (radio and X-ray) \LEM-\HI\ channel combination, which improves over the CCD-\HI\ combination; however, this may be stochastic, as the CCD outperforms the $\mu$cal and there is likely little reason for a $\mu$cal to be able to improve over a CCD for halo mass inference.  The group statistics are more scattered, although this mass bin shows that \HI\ adds little to the inference compared to the X-ray channels.  

We also note that there exists a surprising amount of bias in the \LEM-\HI\ frame of Fig.~\ref{fig:TNG100_M500c}, as shown by the divergence from the 1-to-1 line between $M_{500}\approx 10^{12.5}-10^{13.5}\;\msolar$ and a worse RMSE, which is likely due to the steep decline in dataset size as a function of halo mass. We discuss the sample distribution in \S~\ref{sec:discuss_sample}.

\begin{figure*}
\includegraphics[width=\textwidth]{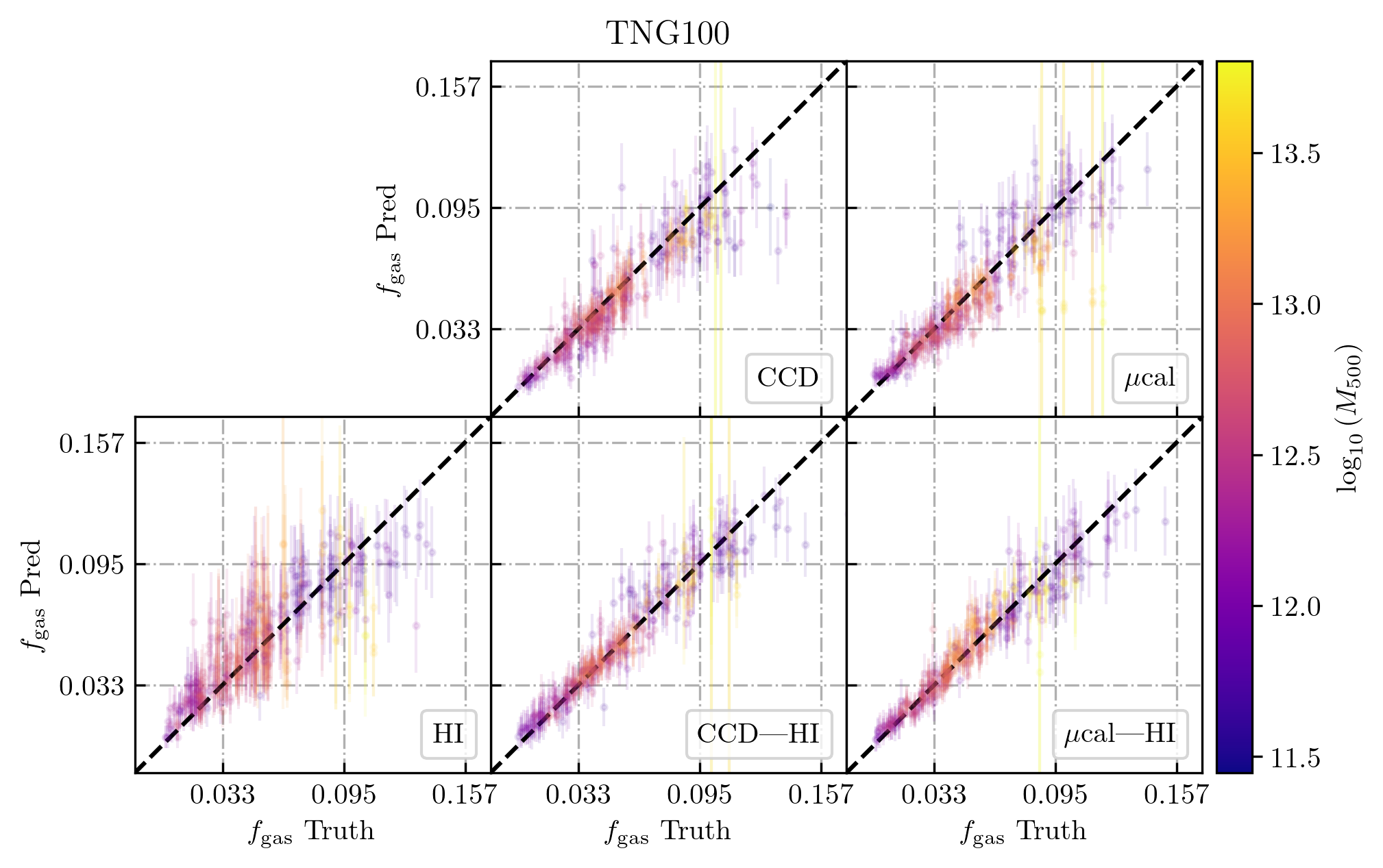}
\caption{Truth-inference plots for $\fgas$ in TNG100, following the same formatting as in Fig.~\ref{fig:TNG100_M500c}, with the same mass-dependent mask to the plotted points.}  
\label{fig:TNG100_f_gas}
\end{figure*}

Figure~\ref{fig:TNG100_f_gas} shows the Truth-Inference results for the gas fraction of the CGM, $f_{\rm gas}$, as defined in Equation~\ref{eq:fgas}. $f_{\rm gas}$ is a non-monotonic function of the halo mass in TNG100, being higher at $L^\star$, dropping at super-$L^\star$ masses, and then increasing again at group masses \citep{nelson2018,davies2020}. Furthermore, there is significant scatter with the halo mass, making inference of this property a test that is not directly dependent on an underlying correlation with $M_{500}$. $\HI$ again performs worst, while the X-ray-only (CCD and \LEM) inferences perform relatively better at higher masses, as detailed in Table \ref{tab:TNG100_M500}.  The additional \LEM\ channels do not consistently improve over CCD for $\fgas$.  The \LEM\ does not consistently outperform CCD, although \LEM-\HI\ performs best overall for $L^\star$ and super-$L^\star$ halos.  Overall, two wavebands (\HI\ and X-ray) tracking different gas phases aid the inference of $f_{\rm gas}$, as also shown by G24.  

\begin{figure*}
\includegraphics[width=\textwidth]{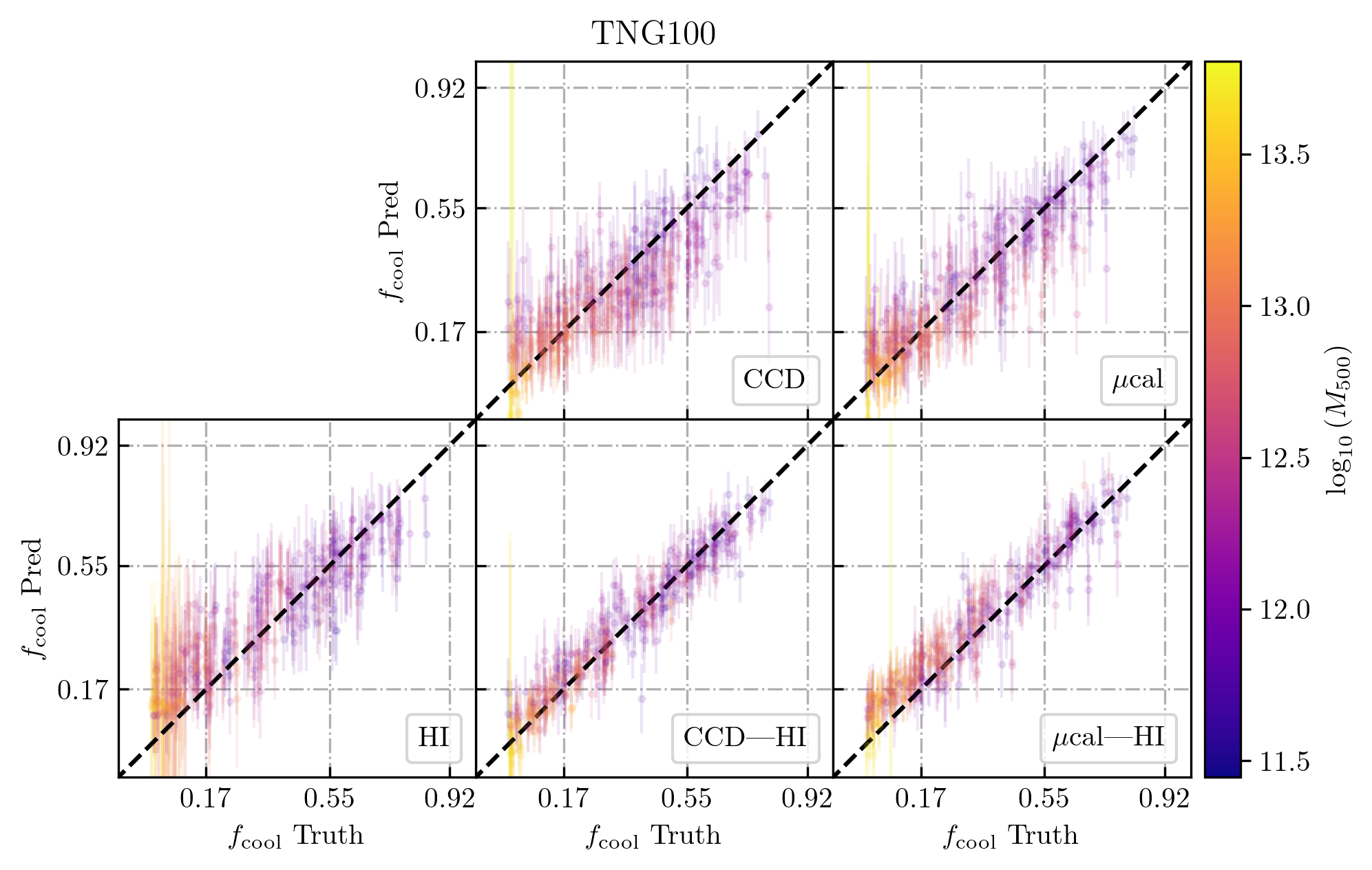}
\caption{Truth-inference plots for $\fcool$ in TNG100, following the same formatting as in Fig.~\ref{fig:TNG100_M500c}, with the same mass-dependent mask to the plotted points.}
\label{fig:TNG100_f_cool}
\end{figure*}

Figure~\ref{fig:TNG100_f_cool} shows the Truth-Inference results for the cool gas fraction of the CGM, $f_{\rm cool}$, as defined by Equation~\ref{eq:fcool}. These maps present another unique inference challenge, as the cool gas fraction is responsible for \HI\ emission, relative to the hot gas, which is traced by X-ray emission. Here, \HI\ alone performs better than X-ray alone, which is expected due to the thermodynamic phases they each trace. However, as with our previous results, the combined CCD-\HI\ and \LEM-\HI\ perform the best.

Table~\ref{tab:TNG100_M500} quantifies the statistical trends.  Looking at the RMSE of $L^\star$ galaxies, \HI\ is 0.096, CCD is 0.120, but when combined, the RMSE is 0.069. \HI\ is more constraining for these galaxies that often exceed $\fcool=0.5$.  For super-$L^\star$ galaxies, RMSE is 0.11 for both \HI\ and CCD, and is 0.058 when combined; hence, both wavebands contribute equally and provide much better inference in combination.  We note that the same situation occurs for $\fgas$ for $L^\star$ galaxies.   

To summarize this section, we infer three properties using five channel combinations from two wavebands.  Halo mass inference is least dependent on multiple channels as long as X-ray is observed, likely because X-ray luminosity is a strong monotonic function of halo mass as opposed to \HI.  With both halo gas fraction, $f_{\rm gas}$, and cool gas fraction, $f_{\rm cool}$, inferences are improved with multi-band inference, but X-ray is most constraining for $\fgas$, while \HI\ is most constraining for $\fcool$. The best performance is \LEM-\HI\ for $L^\star$ galaxies, and there may be additional information within the \LEM\ channels for these low-mass galaxies with weak X-ray emission. There is less consistent improvement for super-$L^\star$ and especially groups using \LEM\ over the CCD; however, we explore metallicity properties where \LEM\ is expected to consistently outperform the CCD in \S\ref{sec:Z_results}.  

\begin{figure*}
\includegraphics[width=\textwidth]{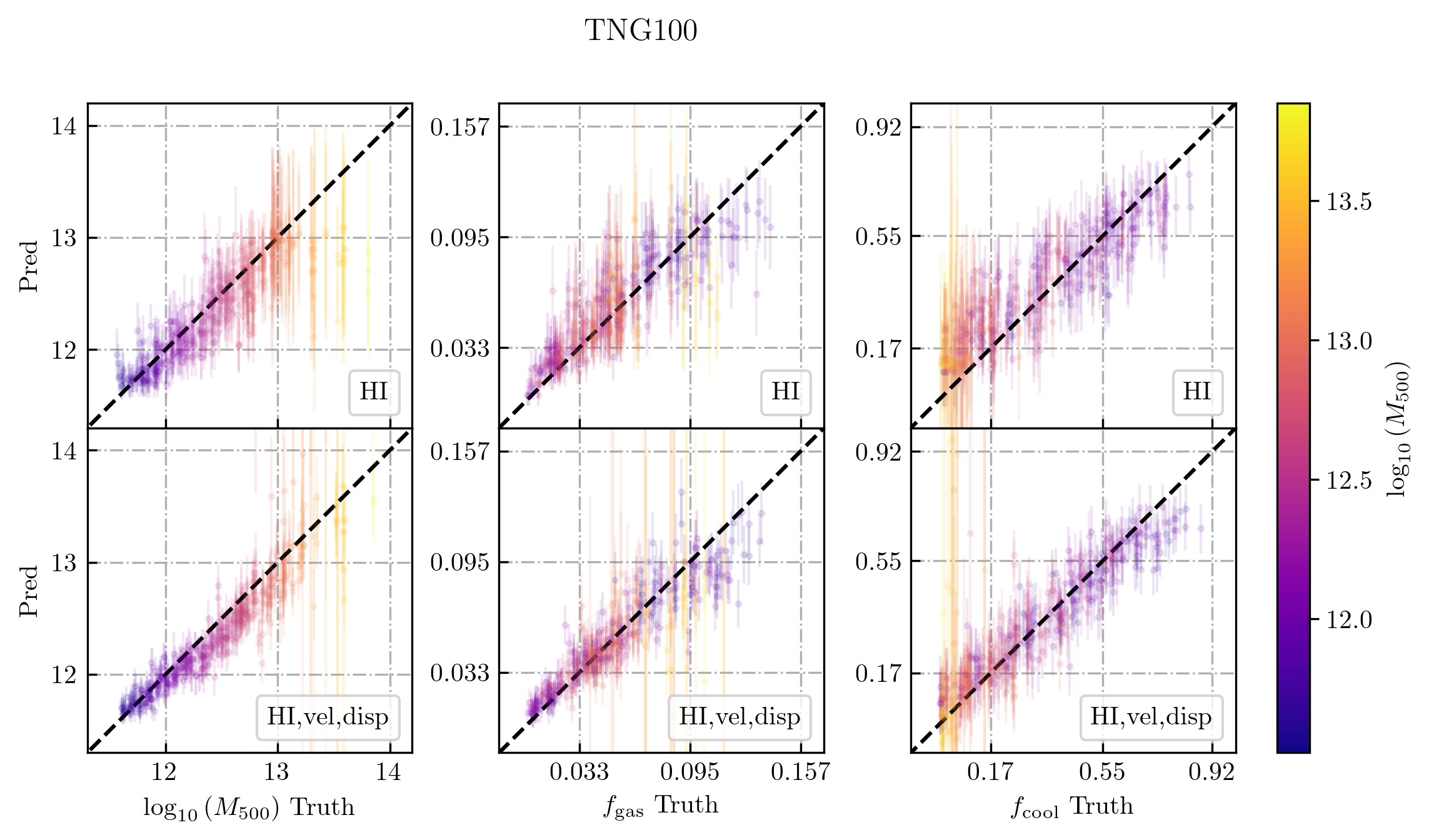}
\caption{Truth-inference plots for \HI\ intensity-only single channel (upper panels) and \HI\ three-channel moment maps including velocity and dispersion maps (lower panels) using the TNG100 simulation.  The three properties displayed are $M_{500}$ (left), $\fgas$ (middle), and $\fcool$ (right).  The upper panels are a repeat of the lower-left panels of Figs. \ref{fig:TNG100_M500c}, \ref{fig:TNG100_f_gas}, and \ref{fig:TNG100_f_cool}.}
\label{fig:TNG100_3_moments}
\end{figure*}

\subsubsection{Inference with Neutral Hydrogen Moment Maps}\label{sec:moment_results}

We now train on the multiple moment maps (see the three columns from the right, of Fig.~\ref{fig:snaps}) to infer the same three properties as in the previous subsection: $M_{500}$, $f_{\rm gas}$, and $f_{\rm cool}$. The first and second moments (velocity and dispersion, respectively) differ from previously used \HI-intensity maps as they contain ``zero'' values where the column density is below $10^{19}\;$cm$^{-2}$. 

Figure~\ref{fig:TNG100_3_moments} shows the results from the \HI-moments in the lower panels, with inferences on $M_{500}$ (left), $f_{\rm gas}$ (middle), and $f_{\rm cool}$ (right). The upper panels in this figure repeat the previous \HI\ results (lower left panels of Figs.~\ref{fig:TNG100_M500c}-\ref{fig:TNG100_f_cool}) for comparison.

The new moment channels show a significant improvement in $M_{500}$ inference for $L^\star$ galaxies, and are competitive with combined X-ray-$\HI$ results (see Table~\ref{tab:TNG100_M500}).  Gas velocities can probe the virial velocity, and therefore the virial mass; however, there is no similar improvement for super-$L^\star$, which may be in part due to bias for this run. However, X-ray imaging is far superior to \HI\ moment maps above the $L^\star$ bin. There is an appreciable improvement in $\fgas$ RMSE, which decreases by about a third across all mass bins, but this is rarely competitive with X-ray or combined channels. $f_{\rm gas}$ also improves across all mass bins with \HI\ moment maps, and outperforms X-ray alone for both $L^\star$ and super-$L^\star$, but does not outperform combined channels. 

In summary, $\HI$ moment maps appear to provide inference for determining halo masses of $L^\star$ galaxies, which is necessary for determining $\fgas$ and $\fcool$ ratios, but it is usually better to complement $\HI$ with X-rays.  

\begin{figure*}
\begin{center}
\includegraphics[width=0.75\textwidth]{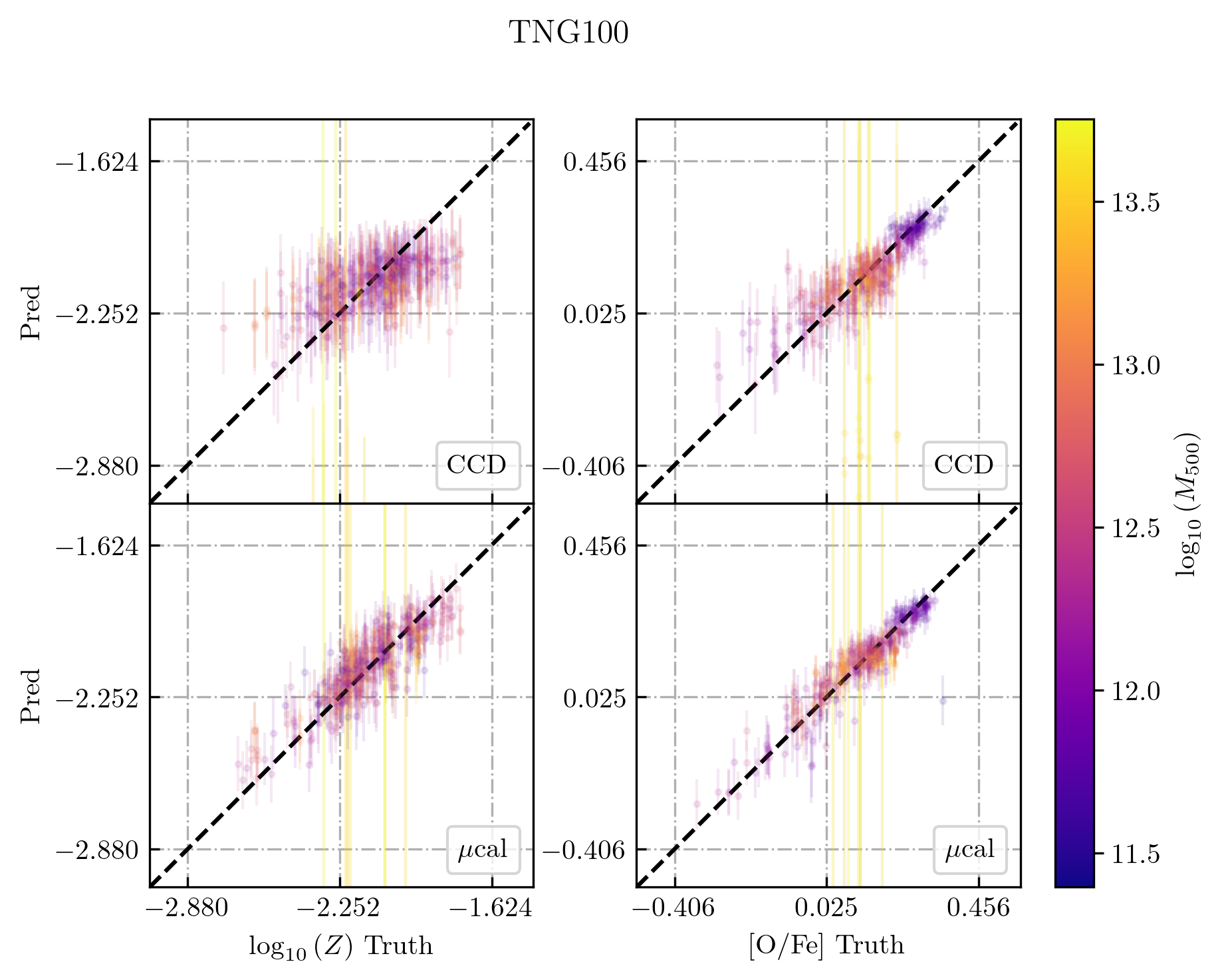}
\end{center}
\caption{Truth-inference plots for metallicity ($\log_{10}(Z)$; right panels) and [O/Fe] abundance ratios (left panels) using the TNG100 simulation. The upper panels show X-ray CCD channels, and the lower panels show micro-calorimeter channels. The same mass-dependent mask applied to Fig. \ref{fig:TNG100_M500c} is used here.}
\label{fig:TNG100_Z_OFe}
\end{figure*}

\subsection{X-ray Metallicity and Abundance Ratios}\label{sec:Z_results}

Figure~\ref{fig:TNG100_Z_OFe} shows the Truth-Inference results for metallicity, $\log(Z)$, of the hot gas (defined by temperatures $T>10^6$ K). This quantity provides a challenging test for CNNs, as this measurement is often independent of $M_{500}$, as discussed by G24. The upper left panel of Fig.~\ref{fig:TNG100_Z_OFe} shows relatively poor inference from using the 4-channel X-ray CCD. G24 also found similarly poor inference in single-channel idealized X-ray maps. However, using the 7-channel \LEM\ setup (upper right panel of Fig.~\ref{fig:TNG100_Z_OFe}) reveals much improved inference for super-$L^\star$ halos (RMSE declines from 0.15 to 0.085) and group halos (RMSE declines from 0.16 to 0.095; see Table~\ref{tab:TNG100_M500}). 

This result indicates that metallicity is measurable in \LEM\ observations that combine three narrow bands centered on metal emission lines ($\OVII$, $\OVIII$, and $\FeXVII$) and four wide bands (0.2-0.4 keV, 0.4-0.6 keV, 0.73-1.1 keV, and 1.43-2.0 keV). Three of these wide bands are specifically tailored to excise strong metal emission and focus on Bremsstrahlung emission dominated by primordial elements.  The multi-channel CNN appears to learn that a combination of these channels is sensitive to measuring metallicity.  Furthermore, the result indicates that $\log(Z)$ is measurable in the hot CGMs of the lowest mass bin $L^\star$ galaxies, given a 1 Msec total integration. The full soft X-ray band ($0.5-2.0$ keV) is dominated by metal-line emission for these galaxies; hence, this result suggests that the CNN is learning to ratio the narrow metal-line channels with the continuum channels.   

We also predict the [O/Fe] abundance ratio of the hot gas, which measures $\alpha$-enhancement of metal-enriched gas.  There is a more obvious correlation of [O/Fe] with halo mass, which decreases from $L^\star$ to super-$L^\star$. However, inference improves with \LEM\, showing the highest increase over CCD for super-$L^\star$ halos (RMSE$=$0.059 for CCD improves to 0.040 for \LEM). This result suggests that the \LEM\ CNN is learning to ratio the ionized oxygen channels ($\OVII$ and $\OVIII$) with the $\FeXVII$ channels to more accurately infer the hot gas [O/Fe].  We argue that this demonstrates the superiority of \LEM\ for measuring $\alpha$-enhancement and aiding in finding the origin of the elements in the CGM. The lowest-mass galaxies have higher [O/Fe], likely owing to more recent outflows associated with late-time supernovae with higher $\alpha$-enhanced nucleosynthetic products \citep[e.g.][]{dave2008}.   

\begin{figure*}
\includegraphics[width=\textwidth]{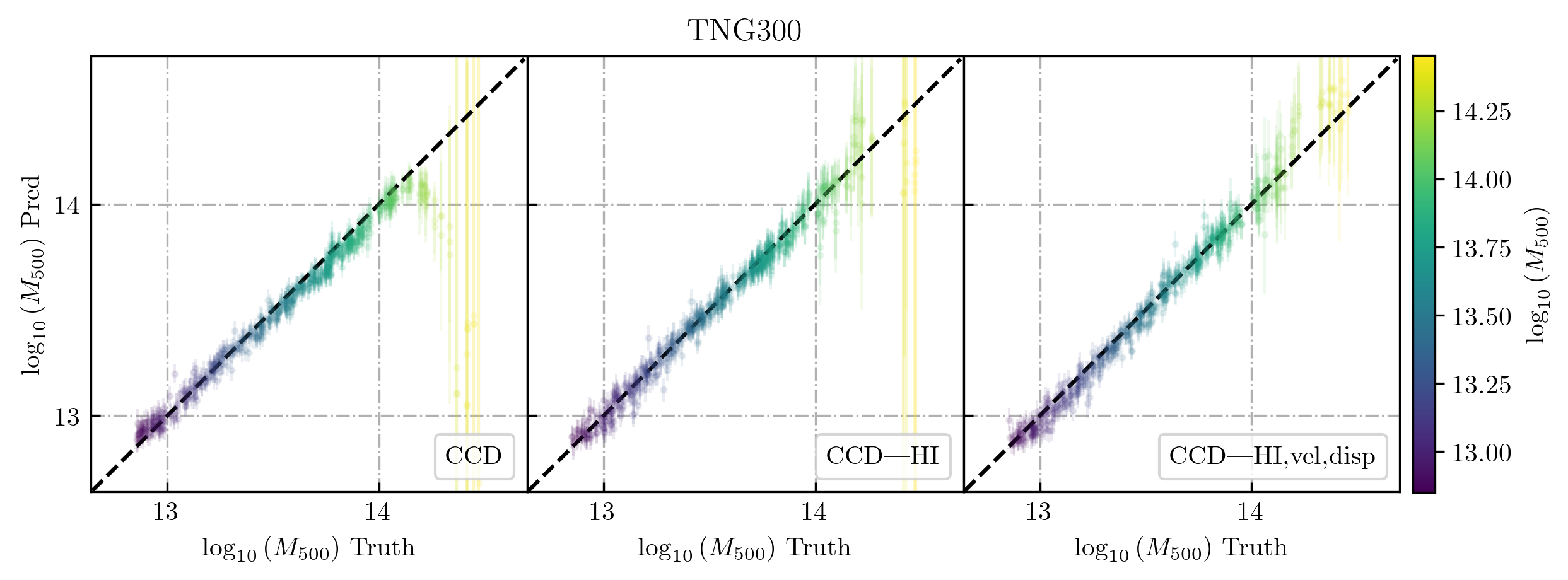}
\includegraphics[width=\textwidth]{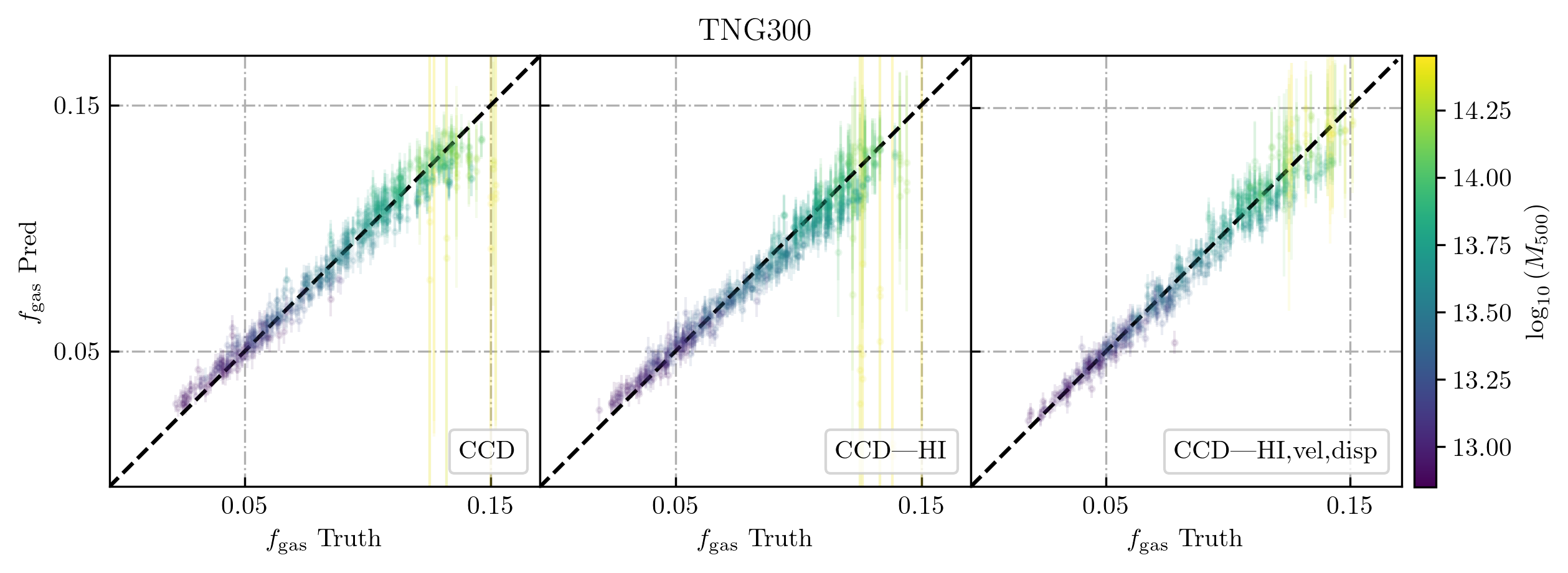}
\caption{TNG300 truth-inference plots for $M_{500}$ (upper panels) and $\fgas$ (lower panels) using X-ray CCD channels (left), CCD-\HI\ intensity (middle), and CCD-\HI\ moment maps (right).  The mass range covers groups and low-mass clusters in the TNG300 volume, and data points are colored by $M_{500}$. A similar mass-dependent mask of data points is applied here as with \ref{fig:TNG100_M500c}; however, the mass range has increased to fit the simulation population.}
\label{fig:TNG300_M500}
\end{figure*}

\begin{center}
\begin{table*}[t]
	\centering
	\caption{The inference statistics for TNG300 results, divided into $M_{500}$ mass bins. Underneath the inferred properties ($M_{500}$ and $\fgas$), the channels used as inputs to the CNN are listed. The results of both properties are shown in Fig. \ref{fig:TNG300_M500}. Bold values indicate the best value in each column for each property.
    } 
    \begin{tabular*}{13.4cm}{|m{2.8cm}|m{0.9cm}|m{0.7cm}|m{0.7cm}|m{0.9cm}|m{0.7cm}|m{0.7cm}|m{0.9cm}|m{0.7cm}|m{0.7cm}|}
	\hline
	\textbf{Channels} & \multicolumn{3}{c|}{\textbf{$<10^{13.35}$}} & \multicolumn{3}{c|}{\textbf{$10^{13.35}-10^{14.0}$}} & \multicolumn{3}{c|}{\textbf{$>10^{14.0}$}} \\
	\cline{2-10} & RMSE & $\rho$ & $\chi^2$ & RMSE & $\rho$ & $\chi^2$ & RMSE & $\rho$ & $\chi^2$ \\
	\hline
    \sechead{$M_{500}$}\\
	\hline
	CCD & \textbf{0.041} & \textbf{0.96} & 0.71 & 0.046 &  0.98 &  \textbf{0.82} & 0.16 & -0.25 & \textbf{1.18} \\
	CCD-\HI & 0.045 &  0.95 & 0.70 & \textbf{0.033} & 0.98 & 0.32 & \textbf{0.096} & 0.78 & 0.28 \\
	CCD-\HI,vel,disp & 0.049 & 0.95 & \textbf{0.85} & 0.041 & 0.98 & 0.63 & 0.099 & \textbf{0.91} & 0.31 \\
	\hline
    \sechead{$\fgas$}\\
	\hline
	CCD & 0.004 & 0.96 & \textbf{1.05} & \textbf{0.005} & 0.96 & 0.88 & \textbf{0.009} &  0.25 & \textbf{0.53} \\
	CCD-\HI & 0.004 & \textbf{0.97} & 0.72 & 0.006 & 0.96 &  0.85 & 0.012 & 0.28 & 0.23 \\
	CCD-\HI,vel,disp & 0.004 & \textbf{0.97} & 0.74 & \textbf{0.005} & 0.96 & \textbf{0.94} & \textbf{0.009} & \textbf{0.67} & 0.48 \\
    \hline
	\end{tabular*}
\label{tab:TNG300_M500_fgas}
\end{table*}
\end{center}

\subsection{Inferring Halo Mass with TNG300}\label{sec:TNG300_M500c}

Inferring group-sized halo properties ($10^{13.0}$--$10^{14.0}\ M_{\odot}$) using TNG100 is difficult due to only having 168 halos available within the small volume. The quantity is further decreased when the full dataset is split for training, validation, and testing. By utilizing the $20\times$ larger TNG300 simulation, we show that increasing the amount of data available, from 168 to 3545 halos in this same mass range, for training, drastically improves inferences for these group-sized halos.

Figure~\ref{fig:TNG300_M500} shows truth-inference plots using 4-channel X-ray CCD, 5-channel CCD-\HI, and 7-channel CCD-\HI\ moment maps for $M_{500}$ (top row) and $\fgas$ (bottom row). In particular, the statistics in Table \ref{tab:TNG300_M500_fgas} demonstrate that training on groups returns much better statistics in the $10^{13}-10^{14}\;\msolar$ range. However, clusters ($>10^{14}\;\msolar$) are poorly inferred due to their low numbers (280 cluster-sized halos in TNG300), in the same problem shifted to higher mass.  

The inconsistent improvement using \HI\ shows that this waveband adds little to inferring the mass of groups and clusters.  For $M_{500}$, the inference remains unimproved for low-mass groups, but is improved for high-mass groups and clusters, which may relate to the ability for \HI\ detected in satellite galaxies to provide another handle on the halo mass.  We show the 3-channel \HI\ moment maps added to X-ray for the first time, and we do not show the results for TNG100 because they did not improve inference.  For $M_{500}$, the moment maps do not improve significantly over \HI\ alone.  For $\fgas$, the CCD-\HI\ moment maps perform the best overall, though the improvement is tiny and within the bounds of run-to-run scatter. Additionally, $\fgas$ is a monotonic function of $M_{500}$ in this mass range, unlike the non-monotonic trends seen in the TNG100 mass range.

\section{Discussion} \label{sec:discuss}

\subsection{Performance and Sample Distribution}\label{sec:discuss_sample}

Our CNN performance declines significantly for the most massive halo bin because it has a smaller sample size in both TNG100 and TNG300, due to the strict volume limit of these simulations. Thus, future investigations should consider using zoom-in simulations to create a more even distribution of samples per halo mass bin. The TNG-Cluster Project \citep{Nelson2024} focuses entirely on more massive clusters, and does not include halos with masses below $10^{14}\ M_{\odot}h^{-1}$.  The CARPoolGP zoom-in simulations \citep{Lee2024} use the IllustrisTNG code to extend \texttt{CAMELS} \citep{camels_2021} with an additional 768 halos of masses $10^{13}-10^{14.5}\ M_{\odot}h^{-1}$. However, the limited \texttt{CAMELS} box-size does not allow for many cluster-sized halos ($M> 10^{14.5}\ M_{\odot}h^{-1}$).

\subsection{Comparison to Previous Networks }\label{sec:discuss_previous}

The CNN framework presented here builds directly on G24, which also applied a CNN to the \texttt{CAMELS} simulations (IllustrisTNG, SIMBA, and Astrid CV sets) to infer CGM properties from combined X-ray and HI maps. Our results showing that multiwavelength observations outperform single-band inference for $f_{\rm gas}$ and $f_{\rm cool}$ are broadly consistent with the trends in G24. However, a direct numerical comparison of performance metrics is not straightforward because of differences in simulation choice, observational setup, and property definitions, which we outline below. We also note a difference in definitions: G24 uses Sub-$L^{\star}$ for halos of masses $\log(M_{\rm halo}/M_{\odot}) \in [11.5, 12]$, $L^{\star}$ for $\log(M_{\rm halo}/M_{\odot}) \in [12, 13]$, and groups for $\log(M_{\rm halo}/M_{\odot}) \in [13, 14.3]$. We use $L^{\star}$ for halos of masses $\log(M_{\rm 500}/M_{\odot}) < 12.3$, Super-$L^{\star}$ for $\log(M_{\rm 500}/M_{\odot}) \in [12.3, 13.0]$, and groups for $\log(M_{\rm 500}/M_{\odot}) > 13.0$. 

The most fundamental difference is the choice of simulation suite. G24 uses the \texttt{CAMELS} CV set, which has a $(25\ h^{-1}{\rm Mpc})^3$ box size that is well-suited to study parameter variation, but too small to contain a statistically significant sample of halos with masses above $10^{13.5}\ M_{\odot}$. This limited G24's group sample to about 190 halos per simulation, and even fewer $10^{14.0}\ M_{\odot}$ objects. By contrast, TNG100 and TNG300 provide both the resolution to study typical $L^{\star}$ halos and the volume needed to sample the group-mass regime, enabling reliable inference for these higher-mass objects. IllustrisTNG is also a well-validated simulation widely benchmarked against CGM and intracluster medium observations, making our results more directly comparable to the broader literature \citep[e.g.][]{nelson2018,Nelson2024}.

The observational setups also differ substantially. G24 used a single broadband soft X-ray channel (0.5-2.0 keV) with eROSITA eRASS:4 \citep{predehl_erosita_2021} observational limits (surface brightness limit $2\times 10^{-13}\ {\rm erg\ s}^{-1} {\rm cm}^{-2} {\rm arcmin}^{-2}$), combined with a single \HI\ column density channel with MHONGOOSE survey limits (\HI\ column density $N_{\rm HI}=10^{19}\ {\rm cm}^{-2}$). Our simulated maps instead use up to seven \LEM\ channels (4 widebands and 3 narrowbands centered on metal emission lines $\OVII$, $\OVIII$, and $\FeXVII$) that include X-ray backgrounds and their subtraction, as well as $\HI$ moment maps encoding velocity and dispersion. We also assume much deeper exposure times (1 Msec for TNG100, 100 ksec for TNG300) compared to G24 {\it eROSITA}-depth limits (eRASS:4 limits of 1 ksec). Because of these differences in simulation, observational depth, and channel configuration, we do not make direct numerical comparisons of combined X-ray and \HI\ performance between the two studies.

However, the qualitative agreement between our results and G24 is reassuring and validates the general framework. Both studies find that X-ray observations powerfully constrain halo mass via the strong monotonic relationship between X-ray luminosity and $M_{\rm 500}/M_{\rm halo}$ while $\HI$ alone performs poorly for this property at higher masses. Both studies confirm that $\HI$ is more informative than X-ray for $f_{\rm cool}$, and that the combination of both wavebands consistently outperforms either alone across $L^{\star}$ and super-$L^{\star}$ halos for $f_{\rm gas}$ and $f_{\rm cool}$ inference.

For CGM metallicity inference, we can make a more direct and physically motivated comparison, as the key driver of improvement here is spectral resolution rather than exposure depth. G24 found poor metallicity inference using single broadband soft X-ray (RMSE$\sim$0.25 dex for halos of mass $\log(M_{\rm halo}/M_{\odot}) \in [12,13]$, corresponding to $L^{\star}$ halos here). Our results using the analogous broadband X-ray CCD configuration are consistent with this conclusion (RMSE$\sim$0.11 dex for $L^{\star}$ halos). Furthermore, decomposing the soft X-ray band into narrow-band channels centered on $\OVII$, $\OVIII$, and $\FeXVII$ substantially improves metallicity inference, reducing the RMSE from 0.15 to 0.085 dex for super-$L^{\star}$ and from 0.16 to 0.095 dex for group halos. Since our CCD and $\mu$cal mocks assume the same exposure time, this improvement isolates spectral resolution as the critical factor for inference of CGM metallicity. 

Beyond the metallicity comparison, our framework extends G24 in two additional ways. First, adding $\HI$ velocity and dispersion moment maps improves column density alone. For inference of the halo mass in $L^{\star}$ halos (noting the difference in the definitions of the halo mass between $M_{\rm 500}$ here and $M_{\rm 200}$ in G24), including velocity and dispersion maps reduces the RMSE from 0.15 ($\HI$ column density alone) to 0.097 ($\HI$ moment maps), making moment maps competitive with the combined results of the column density of X-ray and \HI \ for this mass bin. For $f_{\rm cool}$, moment maps improve the RMSE across all mass bins, from 0.096 to 0.081 for $L^{\star}$ and from 0.11 to 0.083 for super-$L^{\star}$ halos. This improvement indicates that cool gas kinematics more directly trace the cool phase fraction than intensity alone. 

Second, by leveraging the larger box size of TNG300 that included more high-mass halos, we demonstrate that the 100~ksec X-ray CCD exposures can recover group halo masses to high accuracy in the $L^{\star}$ and super -$L^{\star}$ halos (RMSE$\sim$0.041-0.046 dex), a regime where G24 explicitly noted their results were statistically limited.

\section{Summary} \label{sec:summary}

We further develop a Convolutional Neural Network (CNN) based on \citep{gluck_observationally_2024} to assess the effectiveness of existing and future facilities for measuring key properties of dark matter and gas halos via deep imaging observations of their diffuse gas reservoirs. Our purposes in this paper are twofold.   

First, we evaluated the information content using deep learning techniques to guide the effectiveness of deep observations. Long X-ray exposures and deep pointings with \HI\ radio surveys are both extremely expensive observations that require significant resources. Thus, we focus on these facilities' ability to measure key astrophysical quantities and estimate the requirements for future surveys. Our findings using our CNN include: 
\\

\begin{itemize}
    \item Coordinated observations between a soft X-ray CCD and radio telescope mapping 21-cm \HI\ can significantly improve measurements of $\fgas$ for $L^\star$ galaxies and $\fcool$ for $L^\star$ and super-$L^\star$ galaxies. 

    \item Group-scale halo masses ($M_{\rm halo} \geq 10^{13}$~M$_\odot$) can be recovered to high accuracy (RMSE $\sim$0.04~dex) with 100~ksec X-ray CCD exposures, a regime where the larger TNG300 sample provides robust statistics not available with TNG100 or \texttt{CAMELS}.

    \item The addition of \HI\ velocity and dispersion moment maps provides a meaningful improvement over \HI\ intensity mapping alone, especially for L$^\star$ galaxies, where cool gas kinematics directly trace the cool phase fraction. However, \HI\ observations combined with X-ray data consistently outperform \HI\ alone across all halo mass bins, emphasizing the fundamental importance of coordinated multi-wavelength strategies for characterizing global halo properties.

    \item A microcalorimeter X-ray mission with high spectral resolution enables superior measurement of CGM metallicity around L$^\star$ and super-L$^\star$ galaxies, improving over CCD by a factor of 1.75. Clean separation of \OVII, \OVIII, and \FeXVII\ line emissions from the Galactic foreground also enables [O/Fe]-sourced $\alpha$-enhancement measurements with significantly higher accuracy, especially around super-L$^\star$ galaxies.
   
\end{itemize}

Our exploration considers using deep learning to provide a quantifiable evaluation of costly observational strategies for obtaining fundamental physical properties of gaseous halos.  To be clear, DL-based techniques are not necessary to demonstrate that a micro-calorimeter is superior to an X-ray CCD or that 3-channel \HI\ is better than \HI\ intensity.  However, when evaluating a proposed microcalorimeter mission like {\it ExCEED}, we argue that DL-based techniques can guide multi-band synergies that are harder to evaluate with traditional methods.  Our results also highlight the inherent limitations of \HI-only observations for characterizing global gas halo properties, given the vast reservoir of hot gas missed in 21-cm surveys --- a limitation that further motivates coordinated X-ray and radio strategies.

The hot CGM remains one of the most poorly constrained reservoirs in galaxy evolution, and our results point directly toward what is needed to change that. {\it ExCEED}, a microcalorimeter mission concept currently in development, offers the capabilities needed for this challenge. Its 1--2~eV energy resolution and large grasp are precisely what is needed to detect faint CGM emission at large galactocentric radii and cleanly separate \OVII, \OVIII, and \FeXVII\ line emission from the Galactic foreground. The stakes are high: state-of-the-art cosmological simulations currently span orders of magnitude in their predictions for the extent, temperature structure, and brightness of the hot CGM around Milky Way-mass galaxies, and only a mission with {\it ExCEED}'s capabilities can adjudicate between them. Our work delivers the simulation-grounded predictions and discriminating diagnostics that {\it ExCEED} will need to turn its first observations into physical insight, making this a timely and compelling contribution to the science case for the next generation of soft X-ray observatories.

Second, the framework presented here serves as a foundational component for the combined Deep Learning Halo Definer (DLHD) network introduced in Ogle et al. (in prep). While our current results are entirely based on mock data from IllustrisTNG simulations, future work will focus on applying this architecture to real observational datasets. Testing the network on actual 21-cm mappings from surveys alongside existing X-ray observations will evaluate its robustness against true instrumental noise and astrophysical foregrounds. Ultimately, utilizing deep learning techniques to quantify the information content of multi-wavelength data will be essential for guiding survey strategies and justifying the long integration times required by upcoming observational facilities. 

\begin{acknowledgments}

We thank Michelle Ntampaka, John Soltis, Barbara Cantinelli, and the WALLABY Team for valuable feedback and suggestions for the paper. The CAMELS simulations were performed on the supercomputing facilities of the Flatiron Institute, which is supported by the Simons Foundation. The convolutional neural network was programmed and executed on the Alpine high performance computing resource at the University of Colorado Boulder. Alpine is jointly funded by the University of Colorado Boulder, the University of Colorado Anschutz, Colorado State University, and the National Science Foundation (award 2201538). This work is supported by the National Science Foundation (NSF) grants AST 2206055 and 2511137 and the Yale Center for Research Computing facilities and staff.
\end{acknowledgments}

\vspace{5mm}

\bibliography{references}{}
\bibliographystyle{aasjournal}

\end{document}